\documentclass[11pt,a4paper]{article}

\usepackage[top=2.5cm,bottom=2.5cm,left=2.5cm,right=2.5cm]{geometry}
\usepackage{amsmath, amssymb, bm}
\usepackage{graphicx}
\usepackage{url}
\usepackage{subfig}
\usepackage{titlesec}
\usepackage{authblk}
\usepackage{setspace}
\usepackage{lineno}       
\usepackage[labelfont=bf,labelsep=period]{caption}

\usepackage[super]{cite}
\usepackage[colorlinks=true,linkcolor=blue,citecolor=blue,urlcolor=blue]{hyperref}

\titleformat{\section}{\large\bfseries}{\thesection}{1em}{}
\titleformat{\subsection}{\normalsize\bfseries}{\thesubsection}{1em}{}

\title{\textbf{Experimental Observation of Metasurface Absorption Beyond the Rozanov Bound via Temporal Modulation}}

\author[1]{Maoxin Wang}
\affil[1]{College of Physics and Optoelectronic Engineering, Harbin Engineering University, China}
\author[1]{Zeyuan Li}
\author[2]{Mohammad Sajjad Mirmoosa}
\affil[2]{Department of Physics and Mathematics, University of Eastern Finland, P.O.~Box~111, FI-80101 Joensuu, Finland}
\author[1,*]{Xuchen Wang}

\affil[*]{e-mail: xuchen.wang@hrbeu.edu.cn}

\date{}

\begin{document}

\maketitle

\begin{abstract}
\noindent\textbf{Traditional passive, linear time-invariant electromagnetic absorbers are fundamentally constrained by the Rozanov bound, which imposes a strict trade-off between absorption bandwidth and physical thickness. Temporal modulation offers a promising approach to circumventing this fundamental limitation. Here, we experimentally demonstrate that a ``metasurface-based" absorber can surpass the Rozanov bound through modulation of its effective parameters in time. Importantly, the applied modulation profile is periodic and non-binary, acts on the reactive rather than dissipative part of the system, and varies slowly relative to the incident-wave frequency. In particular, we realize this concept in a spatially structured electromagnetic platform using a varactor-loaded mushroom-type metasurface with time-varying surface capacitance. The dynamic response is analyzed using Floquet scattering theory, and the fabricated sample is experimentally characterized in a parallel-plate waveguide. The measured thickness-normalized reflection integral reaches 40.93, more than twice the conventional Rozanov bound of 19.74. These findings open new opportunities for developing ultrathin, broadband, and dynamically reconfigurable electromagnetic absorbers, with potential applications in adaptive electromagnetic shielding and interference suppression.}
\end{abstract}

\section*{Introduction}
The absorption of electromagnetic energy underpins critical technologies across various domains, ranging from radar cross-section (RCS) reduction and stealth technologies to energy harvesting, photovoltaics, and electromagnetic interference (EMI) shielding \cite{watts2012metamaterial, zeng2020electromagnetic}. The demand for compact and versatile solutions continuously drives the need for ultrathin, broadband absorbers. Over the past decades, a myriad of engineered structures have been extensively investigated to maximize wave-matter interaction within subwavelength thicknesses, including classic Salisbury screens \cite{salisbury1952absorbent, fante1988reflection}, Jaumann layers \cite{knott1980performance}, Dallenbach layers \cite{dallenbach1938reflection}, as well as modern metamaterial perfect absorbers \cite{landy2008perfect, ra2015thin} and frequency-selective surfaces (FSS) \cite{munk2005frequency, sun2012broadband}. Despite these significant advancements, all passive linear time-invariant absorbers are fundamentally constrained by the Rozanov limit \cite{rozanov2002ultimate}. This theoretical bound, analogous to the Bode-Fano limit for impedance matching \cite{fano1950theoretical} and mirrored by analogous stiffness-governed integral bounds in acoustics \cite{zhang2026causality}, formalizes a strict physical trade-off: for a given absorber thickness, the integration of the logarithmic reflection coefficient over the entire wavelength spectrum cannot exceed a specific constant determined by the material's static permeability. Although recent dispersion-engineered passive matching networks can push the bandwidth-to-thickness ratio remarkably close to this theoretical ceiling \cite{nayani2025passive}, passive LTI systems remain fundamentally incapable of overcoming this intrinsic bandwidth-thickness constraint.

Surpassing the Rozanov bound requires relaxing at least one of its underlying assumptions---namely passivity, linearity, or causality. For example, active absorbers incorporating non-Foster elements created by negative-impedance converter (NIC) circuits can exceed this limit by breaking passivity~\cite{mou2016design}; however, such active designs inherently face challenges in conditional stability, noise generation, and high-frequency scaling due to the finite gain-bandwidth product of active amplifiers. Alternatively, Wang et al. demonstrated a noncausal active acoustic absorber that exploits a priori information about incident waves through feedforward control, thereby surpassing the conventional causal thickness–bandwidth limit~\cite{wang2024breaking}. Beyond approaches that relax passivity or causality, breaking time-invariance through time-varying electromagnetic systems offers another promising route to overcoming the bandwidth--thickness trade-off of passive LTI absorbers~\cite{caloz2019spacetime, engheta2020metamaterials, galiffi2022photonics, hayran2023using}. By introducing the time dimension as a new degree of freedom, dynamic systems break continuous time-translation symmetry, unlocking exotic wave phenomena such as non-reciprocity~\cite{taravati2020space, zang2019nonreciprocal}, frequency conversion \cite{moussa2023observation}, and parametric amplification \cite{lyubarov2022amplified, asadchy2022parametric}. Recent studies have also demonstrated that temporal switching can extend impedance matching bandwidths beyond static limits for short pulses \cite{shlivinski2018beyond, yang2022broadband}.

In the context of absorption, time-modulated strategies such as phase-switched screens \cite{chambers2005smart} and spread-spectrum metasurfaces \cite{wang2021spread} reduce in-band reflections by redistributing narrowband energy across a wide spectrum, rather than genuinely absorbing broadband incident energy. To achieve true absorption beyond static bounds, time-switched absorbers have been explored \cite{li2020temporal, firestein2022absorption}, showing potential to overcome static bandwidth-thickness constraints, yet the stringent requirement on switching speed poses significant practical challenges. Periodic temporal modulation represents another actively investigated time-varying strategy, often discussed in the context of photonic time crystals (PTCs) and Floquet metamaterials \cite{saha2023photonic, wang2023metasurface, yin2022floquet}. For instance, coherent low-frequency modulation of resistive metasurfaces has been theoretically proposed to achieve perfect absorption at multiple discrete frequencies beyond static bandwidth limits \cite{mostafa2022coherently}, though it remains limited to theoretical analysis under discrete multi-tone excitation rather than continuous broadband spectra. Meanwhile, theoretical studies have demonstrated that periodically modulating the permittivity of a highly lossy medium can enhance broadband absorption \cite{hayran2024beyond}, surpassing the Rozanov bound by exploiting parametric absorption and phase-dependent standing-wave interactions in the modulated medium. Nevertheless, implementing high-speed continuous temporal modulation across an entire ``bulk volumetric medium'' at microwave or optical frequencies presents substantial implementation challenges. To address practical implementations, a very recent experimental study successfully observed broadband super-absorption beyond the LTI limits using a time-varying transmission-line (TL) circuit \cite{ciabattoni2025observation}. While this TL-based experiment provided a crucial proof-of-concept for the underlying physics, it was intrinsically restricted to a one-dimensional guided-wave circuit topology. Moreover, the time-varying behavior in that work was realized through an RF switch toggling between discrete resistive states, which restricts the modulation to binary states and limits the modulation frequency due to the finite switching speed. The realization of time-varying absorption phenomena in a scalable, two-dimensional spatial electromagnetic environment---with flexible modulation control---remains an important challenge.

In this work, we for the first time propose and experimentally demonstrate a periodically time-varying metasurface absorber that surpasses the Rozanov limit. By replacing both impractical volumetric bulk modulations and discrete RF-switching schemes with a varactor-loaded mushroom-type metasurface, the system achieves low-frequency periodic temporal modulation of the surface impedance in a spatial wave environment. Compared with the switch-based resistive modulation employed in Ref. \cite{ciabattoni2025observation}, varactor diodes support operation at much higher modulation frequencies and allow the modulation waveform to be flexibly controlled by an external continuous analog voltage, providing a high degree of freedom in tuning the modulation parameters. Furthermore, owing to the time-derivative constitutive relation of a capacitor, capacitive modulation excites substantially stronger Floquet harmonics than resistive modulation, enabling the metasurface to surpass the Rozanov limit with a significantly weaker modulation depth. A generalized circuit-based Floquet framework is employed to analyze the multi-harmonic scattering response and optimize the modulation parameters. The fabricated prototype, validated in a parallel-plate waveguide system, achieves an experimental thickness-normalized Rozanov integral of $40.93$, substantially exceeding the theoretical static limit of $19.74$.

\section*{Results}

\subsection*{Analytical Modeling of Time-Varying Metasurface Absorbers}
\begin{figure*}[!t]
    \centering
    \includegraphics[width=\textwidth]{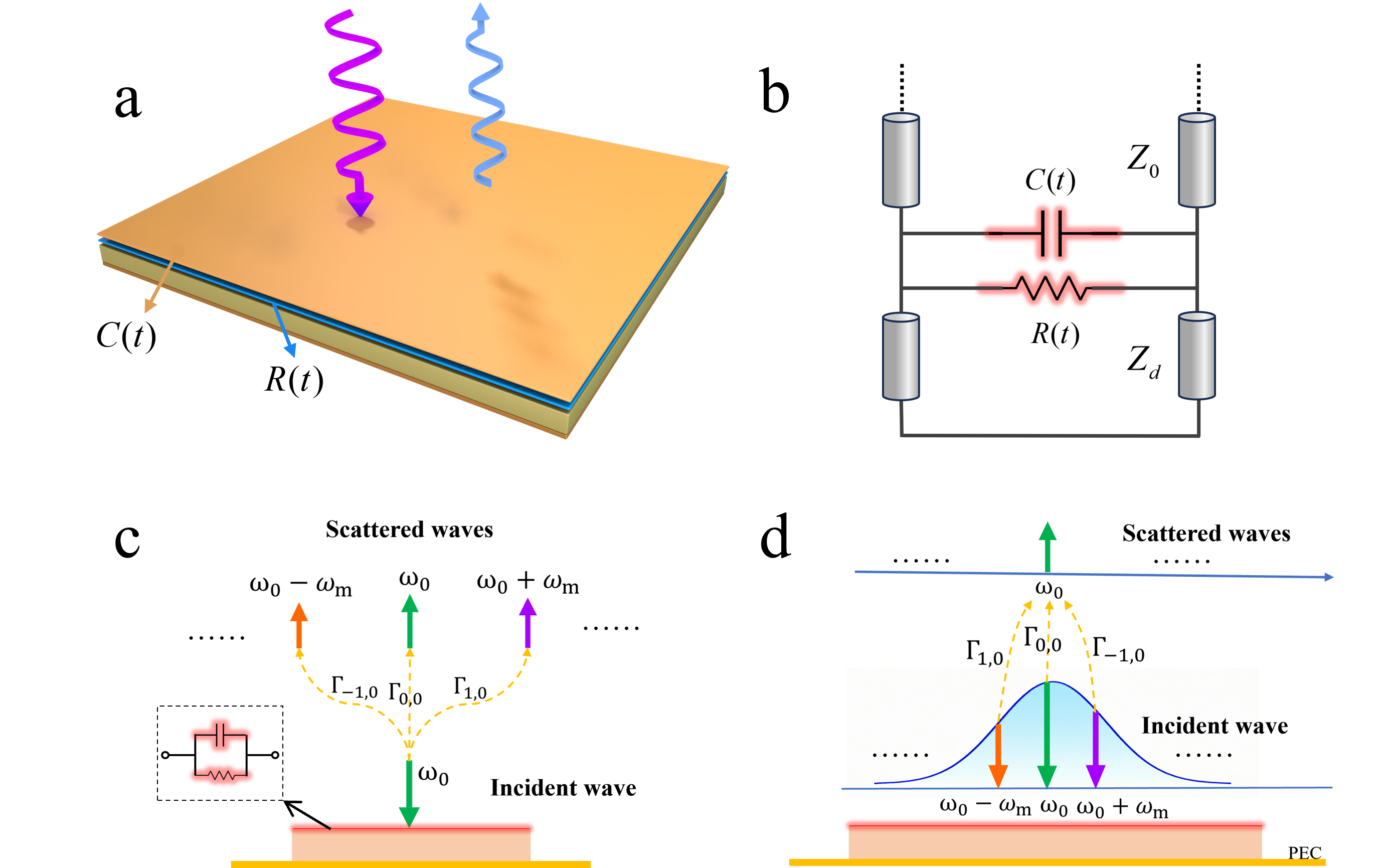}
    \caption{\textbf{Schematic, equivalent transmission-line model, and physical mechanism of the time-varying metasurface absorber.} \textbf{a}, A 2D periodic impedance sheet on a grounded dielectric substrate under normal plane-wave incidence. \textbf{b}, The corresponding transmission-line equivalent circuit model with time-varying surface capacitance $C(t)$ and resistance $R(t)$. \textbf{c}, Single-frequency excitation: the incident fundamental wave at $\omega_0$ is scattered into discrete harmonics $\omega_0 + n\omega_{\mathrm{m}}$ ($n = \pm 1, \pm 2, \ldots$). \textbf{d}, Broadband pulse excitation: different incident frequency components within the pulse spectrum are frequency-converted through distinct harmonic channels and coherently superimposed at the same observation frequency to produce tailored destructive interference.}
    \label{fig:1}
\end{figure*}

We consider the scattering scenario depicted in Fig.~\ref{fig:1}a, where a normally incident plane wave illuminates an electrically thin homogeneous impedance sheet mounted on a grounded dielectric substrate. As illustated by the transmission-line equivalent circuit in Fig.~\ref{fig:1}b, the sheet is modeled by a parallel combination of a time-varying resistance $R(t)$ and a time-varying capacitance $C(t)$, with a common modulation period $T = 2\pi/\omega_{\mathrm{m}}$ where $\omega_{\rm m}$ is the modulation frequency. 
The metal-backed substrate is modeled as a section of a transmission line with length equal to the substrate thickness $d$, characteristic impedance $Z_{\rm d}$, and propagation constant $\beta_{\mathrm{d}}$. The semi-infinite transmission line on the top side represents free space with characteristic impedance $\eta_0$. 

In a conventional time-invariant circuit, voltage and current at each frequency are related by a scalar impedance or admittance, independently of other frequencies. Periodic temporal modulation instead couples different frequencies: an incident wave at $\omega_0$ generates Floquet harmonics at $\omega_n=\omega_0+n\omega_{\mathrm m}$, as illustrated in Fig.~\ref{fig:1}c. Consequently, the current at one harmonic can depend on the voltages at other harmonics, requiring a matrix rather than a scalar circuit description.
Taking the time-varying capacitance as an example, we expand the capacitance and voltage into Fourier series,
\begin{equation}
\begin{aligned}
 C(t)&=\sum_{m=-\infty}^{+\infty} c_m e^{jm\omega_{\mathrm m}t},\\
 v_{\mathrm C}(t)&=\sum_{k=-\infty}^{+\infty}v_k^{\mathrm C}e^{j\omega_k t},
\end{aligned}
\label{eq:fourier}
\end{equation}
where $c_m$ and $v_k^{\mathrm C}$ denote the Fourier coefficient of the capacitance and the voltage amplitude at the $k$th harmonic, respectively. Substituting these expansions into the capacitor relation $i_{\mathrm C}(t)=d[C(t)v_{\mathrm C}(t)]/dt$ and collecting terms at each harmonic gives
\begin{equation}
 i_n^{\mathrm C}=j\omega_n\sum_{k=-\infty}^{+\infty}c_{n-k}v_k^{\mathrm C}.
\label{eq:harmonic_current}
\end{equation}
By truncating the harmonics to $-N\le n,k\le N$, Eq.~\eqref{eq:harmonic_current} becomes
\begin{equation}
 \mathbf i_{\mathrm C}=\mathbf Y_{\mathrm C}\mathbf v_{\mathrm C},
 \qquad [\mathbf Y_{\mathrm C}]_{n,k}=j\omega_n c_{n-k},
\label{eq:cap_matrix}
\end{equation}
where $\mathbf Y_{\mathrm C}$ is a $(2N+1)\times(2N+1)$ harmonic admittance matrix. Its diagonal elements give the response at each harmonic, while its off-diagonal elements describe coupling between different harmonics. In the time-invariant limit, the off-diagonal elements vanish, recovering the uncoupled response of a conventional capacitor.

Similarly, the time-varying resistance is represented by a harmonic admittance matrix $\mathbf Y_{\mathrm R}$, whereas the time-invariant grounded substrate contributes only a diagonal admittance matrix $\mathbf Y_{\mathrm{sub}}$. The derivations of these two matrices are provided in Supplementary Note~1. Their contributions combine to give the total input admittance matrix,
\begin{equation}
 \mathbf Y_{\mathrm{in}}=\mathbf Y_{\mathrm C}+\mathbf Y_{\mathrm R}+\mathbf Y_{\mathrm{sub}}.
\label{eq:input_admittance}
\end{equation}

Enforcing the electromagnetic boundary conditions at the metasurface interface yields the generalized relation between the incident and reflected harmonic voltage vectors,
\begin{equation}
 \mathbf v_{\mathrm{ref}}=\bm\Gamma\mathbf v_{\mathrm{inc}},
 \qquad
 \bm\Gamma=(\mathbf Y_0+\mathbf Y_{\mathrm{in}})^{-1}
 (\mathbf Y_0-\mathbf Y_{\mathrm{in}}),
\label{eq:gamma}
\end{equation}
where $\mathbf Y_0$ is the diagonal free-space admittance matrix with elements $1/\eta_0$. The matrix element $\Gamma_{n,k}$ represents the complex conversion coefficient from an incident harmonic at $\omega_k$ to a reflected harmonic at $\omega_n$. For monochromatic excitation at $\omega_0$, $\Gamma_{n,0}$ therefore specifies the amplitude and phase of the reflected $n$th Floquet harmonic. This matrix formulation provides the basis for analyzing the broadband pulse response discussed in the following section.

\subsection*{Response to Broadband Excitation and Rozanov Limit Bypassing}
The generalized reflection coefficient matrix $\bm{\Gamma}$ derived above characterizes the system response to a single-frequency excitation (see Fig.~\ref{fig:1}c). In this case, under monochromatic illumination at $\omega_0$, the time-varying metasurface scatters the incident wave into a series of discrete harmonics, each weighted by the corresponding element of $\bm{\Gamma}$.

For a broadband incident signal (e.g., a Gaussian pulse), however, the incident wave comprises a continuum of frequency components that interact with the modulated metasurface simultaneously, as illustrated in Fig.~\ref{fig:1}d. In this case, the periodic modulation acts as a frequency mixer: different incident frequencies can be routed into the same observation frequency through different harmonic channels. Specifically, at an observation frequency $\omega$, the reflected field receives contributions not only from the directly reflected component of the incident wave at $\omega$ (i.e., the $n=0$ channel), but also from frequency-converted components of the incident waves at $\omega - \omega_{\mathrm{m}}$ (upconverted through $n=+1$) and $\omega + \omega_{\mathrm{m}}$ (downconverted through $n=-1$), as well as higher-order shifts.

Assuming the incident broadband signal has a frequency spectrum $V_{\mathrm{inc}}(\omega)$, the total reflected spectrum $V_{\mathrm{ref}}(\omega)$ at a specific observation frequency $\omega$ is thus calculated by summing the contributions from all relevant incident frequencies shifted by the modulation harmonics:
\begin{equation}
    \label{eq:broadband_sum}
    V_{\mathrm{ref}}(\omega) = \sum_{n=-N}^{+N} \Gamma_{n,0}(\omega - n\omega_{\mathrm{m}}) \cdot V_{\mathrm{inc}}(\omega - n\omega_{\mathrm{m}}).
\end{equation}
Here, $\Gamma_{n,0}(\omega - n\omega_{\mathrm{m}})$ is the conversion coefficient for the $n$-th harmonic order, describing the scattering of the incident wave at frequency $\omega' = \omega - n\omega_{\mathrm{m}}$ into the observation frequency $\omega$ (since the $+n$-th channel shifts $\omega'$ to $\omega' + n\omega_{\mathrm{m}} = \omega$). The truncation order $N$ must be chosen sufficiently large to ensure that the harmonic summation converges. This spectral summation indicates that the final reflection spectrum is determined by the joint influence of the incident pulse profile and the harmonic scattering characteristics of the time-varying system. It is worth stressing that Eq.~\eqref{eq:broadband_sum} represents a coherent summation of the complex field amplitudes, rather than a sum of their magnitudes or powers. Consequently, both the amplitudes and the relative phases of the individual contributions jointly determine the final reflection spectrum.

To quantitatively describe this broadband response, the effective reflection coefficient for the broadband pulse is defined in the frequency domain as:
\begin{equation}
    \label{eq:eff_gamma}
    \Gamma_{\mathrm{eff}}(\omega) = \frac{V_{\mathrm{ref}}(\omega)}{V_{\mathrm{inc}}(\omega)}.
\end{equation}
Unlike the conventional reflection coefficient of a linear time-invariant (LTI) system, $\Gamma_{\mathrm{eff}}(\omega)$ depends not only on the intrinsic properties of the metasurface, but also on the spectral amplitude and phase distribution of the incident waveform. It therefore characterizes the response to a specific broadband excitation, rather than the inherent single-frequency scattering coefficient of the structure itself.

In this work, the periodically time-varying parameters of the thin sheet (i.e., the resistance and the capacitance) are modulated in a unified form as:
\begin{equation}
    \label{eq:para_mod}
    X(t) = X_0 \left[1 + M_X \cos(\omega_{\mathrm{m}} t + \phi)\right].
\end{equation}
In this expression, $X(t)$ denotes the generic time-varying parameter, which takes the form of $R(t)$ for resistance modulation or $C(t)$ for capacitance modulation; $X_0$ is the corresponding static value ($R_0$ or $C_0$), $M_X$ is the modulation depth specific to each parameter (i.e., $M_R$ for resistance and $M_C$ for capacitance), $\phi$ is the initial phase, and $\omega_{\mathrm{m}} = 2\pi f_{\mathrm{m}}$ is the angular frequency of modulation. Through temporal modulation, incident pulse components at frequencies $\omega-n\omega_{\mathrm{m}}$ ($n\neq0$) are converted to the observation frequency $\omega$, where they coherently superimpose with the directly reflected component ($n=0$). By optimizing the modulation frequency $f_{\mathrm{m}}$, depth $M_X$, and phase $\phi$ relative to the incident pulse, the complex frequency-conversion coefficients can be tailored to promote destructive interference among these contributions, suppressing the reflected spectrum $V_{\mathrm{ref}}(\omega)$ over the operating band. For the prescribed incident pulse, this suppression reduces the input-dependent effective reflection coefficient and can increase the associated Rozanov integral beyond the classical bound for passive LTI absorbers of the same thickness and static permeability. This is possible because temporal modulation introduces coupling between different frequencies, placing the effective reflection response outside the scalar LTI framework underlying the classical bound, while preserving causality.


By evaluating the Rozanov integral over the spectral bandwidth of the incident pulse, we can directly compare the absorption performance of the time-varying system with that of the static case. The thickness-normalized Rozanov integral is defined as:
\begin{equation}
    \label{eq:rozanov_limit}
    \mathcal{I} = \frac{1}{d} \left| \int_{0}^{\infty} \ln |\Gamma_{\mathrm{eff}}(\lambda)| \, d\lambda \right| \leq 2\pi^2 \mu_{\mathrm{s}}.
\end{equation}
For conventional non-magnetic LTI absorbers ($\mu_{\mathrm{s}} = 1$), this bound is strictly $2\pi^2 \approx 19.74$. A non-magnetic system whose Rozanov integral exceeds this bound, namely $\mathcal{I} > 2\pi^2$, is considered to surpass the Rozanov limit.

We further validated the theoretical framework in MATLAB/Simulink, in which only the capacitance is periodically time-modulated. The simulation results show good agreement with those calculated from the theoretical framework. Furthermore, the simulation results confirm the feasibility of surpassing the Rozanov limit via temporal modulation (see Supplementary Note~2 and Supplementary Fig.~1).

To analyze the influence of the modulation parameters and compare the capacitive and resistive modulation approaches, we next examine the parameters of the modulation function given in Eq.~\eqref{eq:para_mod}. The scattering characteristics of the time-varying system are primarily governed by three modulation parameters: the modulation depth $M_X$, the relative phase $\phi$ between the incident pulse and the modulation function, and the modulation frequency $f_{\mathrm{m}}$.

\begin{figure}[!htbp]
    \centering
    \includegraphics[width=1\columnwidth]{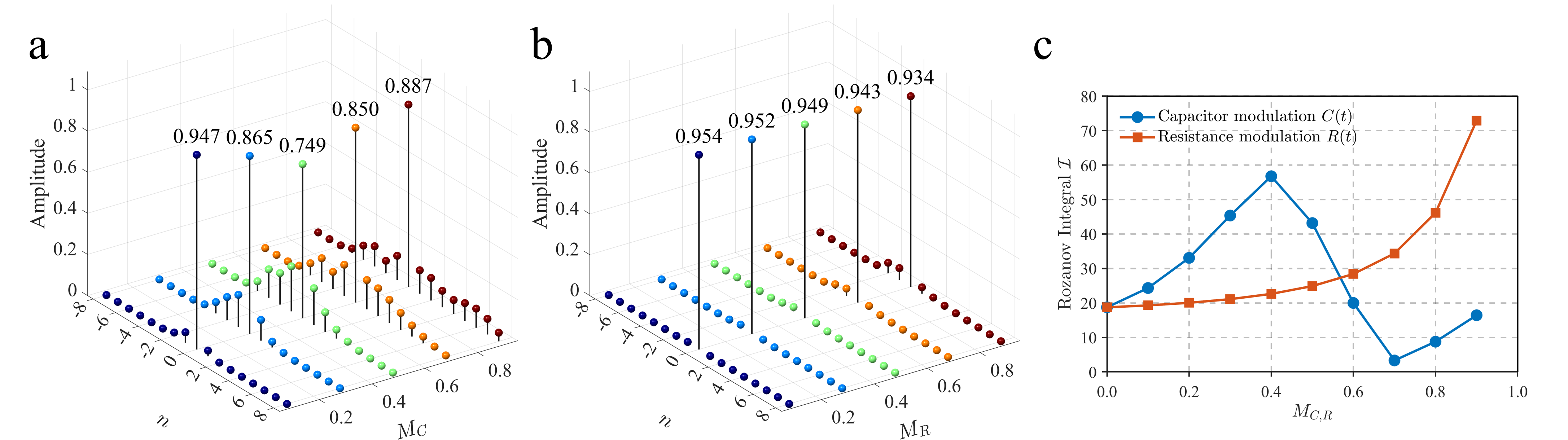}
    \caption{\textbf{Harmonic generation comparison and Rozanov integral analysis under capacitive and resistive modulations.} \textbf{a,b}, Comparison of harmonic generation at different modulation depths ($f_{\mathrm{in}} = 1$~GHz, $f_{\mathrm{m}} = 100$~MHz, $d = 4$~mm) for (\textbf{a}) capacitive modulation ($C_0 = 8$~pF, $M_R = 0$, $R_0 = 300\,\Omega$) and (\textbf{b}) resistive modulation ($R_0 = 300\,\Omega$, $M_C = 0$, $C_0 = 8$~pF). \textbf{c}, Thickness-normalized Rozanov integral evaluated over the 0.5--1.5~GHz frequency range as a function of modulation depth (substrate: $\varepsilon_{\mathrm{r}} = 4.4$, thickness $4$~mm; incident Gaussian pulse consistent with that in Supplementary Note~2; capacitive modulation: $C_0 = 8$~pF, $R_0 = 300\,\Omega$, $f_{\mathrm{m}} = 146$~MHz, $\phi = 1.85\pi$; resistive modulation: $R_0 = 300\,\Omega$, $C_0 = 8$~pF, $f_{\mathrm{m}} = 146$~MHz, $\phi = 0.9\pi$).}
    \label{fig:harmonic_and_rozanov}
\end{figure}

Specifically, the modulation depth $M_X$ governs the coupling strength of the frequency-mixing process. A larger $M_X$ enhances the spectral magnitude of the generated harmonics and extends the harmonic cascade to higher orders. The relative phase $\phi$ determines the timing at which the incident pulse interacts with the temporal modulation cycle, serving as a coherent phase controller. It shifts the relative phase of the generated Floquet harmonics without altering their spectral amplitudes, which is essential for establishing the precise destructive interference needed to suppress the reflected wave. The modulation frequency $f_{\mathrm{m}}$ determines the spectral spacing between adjacent harmonics, directly defining the density of Floquet channels on the frequency axis.

To compare the harmonic generation capability of the two modulation approaches, we examine the time-varying capacitance and time-varying resistance cases within the transmission-line equivalent circuit model of Fig.~\ref{fig:1}b, both modulated following the periodic function form of Eq.~\eqref{eq:para_mod} with the same modulation frequency $f_{\mathrm{m}}$. As shown in Fig.~\ref{fig:harmonic_and_rozanov}a,b, under the same modulation depth ($M_R = M_C = M$), the time-modulated capacitance exhibits significantly stronger harmonic excitation, generating a wider cascade of Floquet harmonics with substantially larger magnitudes than the time-modulated resistance. This difference originates from the mathematical nature of the respective constitutive relations. The capacitive current, $i_{\mathrm{c}}(t) = \frac{d}{dt}[C(t)v_{\mathrm{c}}(t)]$, contains a time-derivative operator that introduces a frequency multiplier $\omega_n = \omega_0 + n\omega_{\mathrm{m}}$, enhancing the coupling to higher-order harmonics. In contrast, the resistive modulation, governed by the purely algebraic relation $v(t) = R(t)i(t)$, lacks this derivative amplification, causing the harmonic amplitudes to decay rapidly with increasing order. This analysis indicates that capacitive modulation has a stronger ability to reshape the broadband pulse spectrum than resistive modulation. Equivalently, to achieve a comparable level of spectral redistribution, capacitive modulation requires a smaller modulation depth than resistive modulation, suggesting that capacitive modulation can achieve enhanced absorption performance with a smaller modulation depth.

We further investigate how the Rozanov integral varies with modulation depth for both capacitive and resistive modulations. Before conducting this comparison, we first select appropriate modulation phase and frequency, as the effective reflection coefficient under time modulation can exceed unity depending on the pulse spectral profile and modulation parameters, which leads to artificially low integral values that are not meaningful for comparison. By fixing suitable values of modulation phase and frequency, we can suppress these undesirable solutions, ensuring a meaningful comparison. The Rozanov integral is then computed as a function of modulation depth $M_{C,R}$, with the results shown in Fig.~\ref{fig:harmonic_and_rozanov}c.

With the modulation frequency and phase properly selected, the Rozanov integral initially increases with modulation depth $M_{C,R}$ for both cases, as the generated harmonics become progressively stronger. The chosen phase and frequency promote coherent destructive interference between the Floquet harmonics and the fundamental response at the same frequencies. For capacitive modulation, however, when $M_C > 0.6$, the harmonic components become excessively strong, causing the effective reflection coefficient to exceed unity at certain frequencies. This results in a sharp decline of the Rozanov integral. In contrast, the resistive modulation produces weaker harmonics throughout, and its Rozanov integral increases monotonically with $M_R$. This comparison demonstrates that capacitive modulation can generate richer harmonic components at a significantly smaller modulation depth, thereby surpassing the Rozanov limit more readily, which is consistent with its stronger spectral redistribution capability discussed above.

\subsection*{Metasurface  Design and Experimental Setup}
\begin{figure}[!htbp]
    \centering
    \includegraphics[width=0.8\columnwidth]{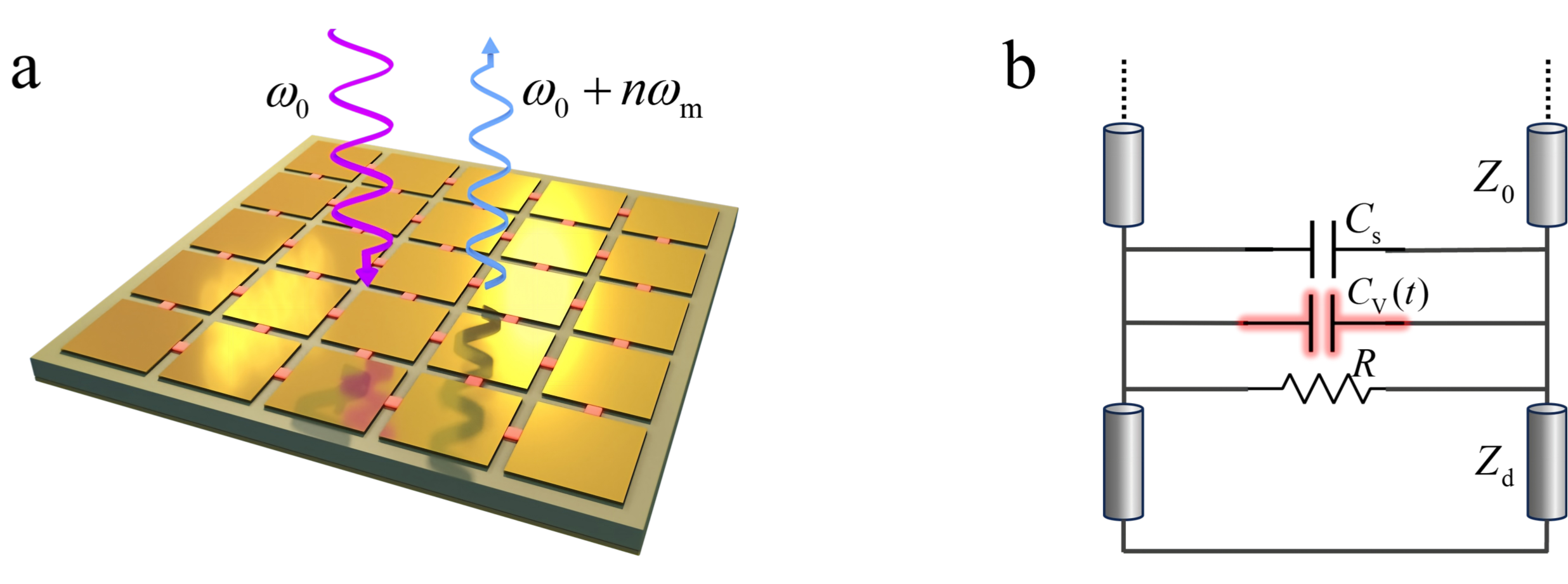}
    \caption{\textbf{Lumped-loaded metasurface array and its transmission-line equivalent circuit.} \textbf{a}, Schematic of the periodic mushroom-type metasurface array with lumped elements bridged across the gaps between adjacent metallic patches. \textbf{b}, Corresponding transmission-line equivalent circuit.}
    \label{fig:unit_cell_model}
\end{figure}

To experimentally validate the proposed time-varying metasurface absorber, we design a physical prototype based on a mushroom-type high-impedance surface (HIS) \cite{sievenpiper1999high_book,sievenpiper2003two}. As illustrated in Fig.~\ref{fig:unit_cell_model}a, lumped elements are bridged across the gaps between adjacent metallic patches in the periodic array, and the corresponding transmission-line equivalent circuit is depicted in Fig.~\ref{fig:unit_cell_model}b. The surface impedance of the metasurface consists of the intrinsic structural capacitance $C_{\mathrm{s}}$ between neighbouring patch edges, the time-varying junction capacitance $C_{\mathrm{v}}(t)$ provided by the reverse-biased varactor diode, and the chip resistor $R_0$ that introduces the ohmic dissipation, all connected in parallel across the patch gaps. Notably, the metasurface must operate under small-signal excitation, where the voltage induced by the incident wave is much smaller than the external direct-current (DC) bias and alternating-current (AC) modulation voltages. This ensures that $C_{\mathrm{v}}(t)$ is driven solely by the external modulation signal rather than being self-modulated by the incident wave. Consequently, the structure can be modeled as a linear time-varying (LTV) system, allowing the Floquet scattering theory derived above to be validly applied to analyze this scenario.

\begin{figure*}[!t]
    \centering
    \includegraphics[width=\textwidth]{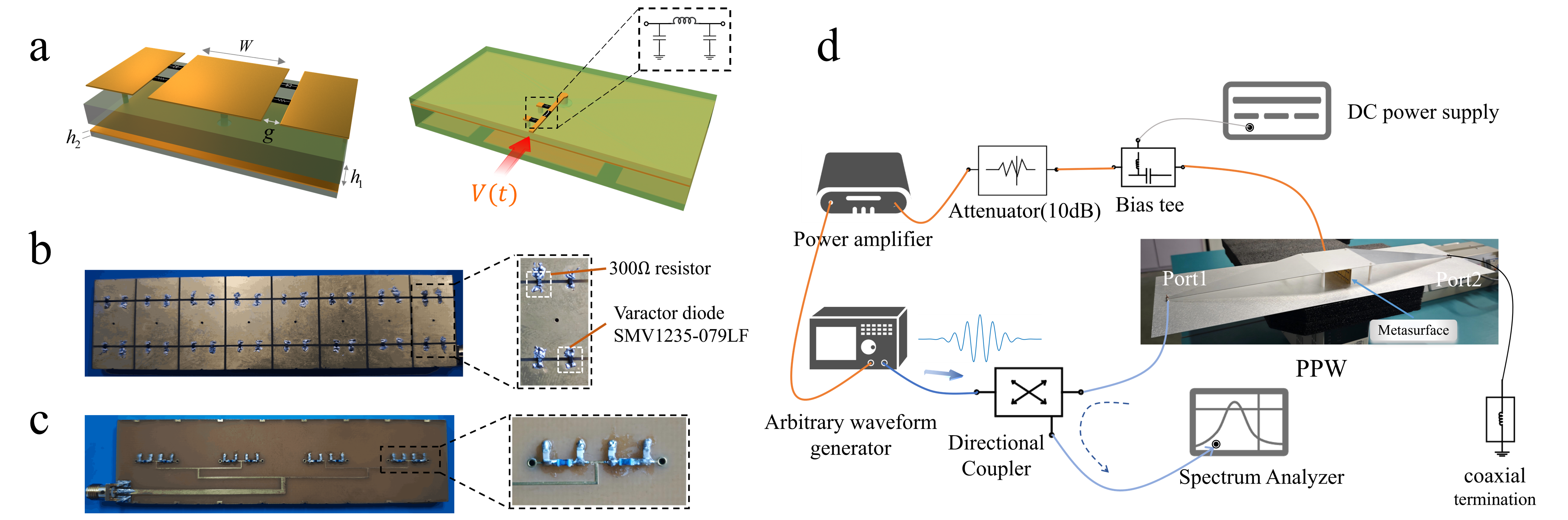}
    \caption{\textbf{Metasurface architecture, fabricated prototype, and parallel-plate waveguide test system.} \textbf{a}, 3D unit-cell geometry of the varactor-loaded mushroom metasurface, showing the top-layer absorbing structure and the bottom-layer biasing network. \textbf{b}, Top-view photograph of the fabricated 1D metasurface array with surface-mount varactor diodes and chip resistors. \textbf{c}, Bottom-view photograph displaying the microstrip feed network integrated with third-order Chebyshev low-pass filters. \textbf{d}, Schematic diagram of the experimental measurement setup configured around the parallel-plate waveguide (PPW), integrating the synchronized pulse generation, continuous modulation pump, and reflected spectral detection.}
    \label{fig:prototype_setup}
\end{figure*}

The structural design and multilayer architecture of the proposed metasurface are illustrated  in Fig.~\ref{fig:prototype_setup}a. The unit cell features a square metallic patch with a width of $W = 15.2$~mm, separated by a gap of $g = 0.6$~mm from adjacent patches. A multilayer printed circuit board (PCB) architecture is employed to distribute the modulation voltage while maintaining strict isolation for the test pulse signals. As shown in the top-perspective view in Fig.~\ref{fig:prototype_setup}a, the structure consists of a top substrate ($h_1 = 3$~mm) for the absorption layer and a bottom substrate ($h_2 = 0.8$~mm) for the biasing network, both utilizing FR-4 material. A continuous ground plane separates these two layers, with holes that allow the metallic vias to pass through. To reveal the bottom-layer biasing mechanism, the bottom-perspective schematic in Fig.~\ref{fig:prototype_setup}a (right) displays the layout of the microstrip traces and the lumped-element filtering network. This integrated filter network ($\pi$-circuit) is specifically designed to ensure effective isolation between the test pulse signals and the modulation signals.

The fabricated 1D prototype is shown in Fig.~\ref{fig:prototype_setup}b,c. Fig.~\ref{fig:prototype_setup}b displays the top view of the metasurface array, where surface-mount varactor diodes (SMV1235-079LF, Skyworks, Inc.) and chip resistors (AT0603BRD07300RL, $300\,\Omega$) are soldered across the gaps between the patches. Fig.~\ref{fig:prototype_setup}c presents the bottom view, displaying the actual microstrip feed network and the integrated third-order lumped-element $\pi$-type low-pass filters. The filters are designed based on a Chebyshev response, with the microstrip trace width set to $0.3$~mm corresponding to a characteristic impedance of $87\,\Omega$. This bottom network distributes the DC bias and low-frequency modulation signals to the top active elements, while the low-pass filters present a high impedance to the higher-frequency test pulse signals, effectively suppressing potential leakage into the external modulation source. The overall dimensions of the fabricated board are $126.4\,\mathrm{mm} \times 31.6\,\mathrm{mm}$.

Rather than fabricating a large-scale two-dimensional (2D) array---which requires extensive active components and a complex biasing network---we configure the prototype for evaluation within a parallel-plate waveguide (PPW) \cite{feng2025investigation, mou2016design} (Fig.~\ref{fig:prototype_setup}d). Since the PPW supports a fundamental transverse electromagnetic (TEM) mode, its top and bottom metallic plates act as electric walls, providing periodic boundary conditions along the transverse direction. Thus, a finite one-dimensional (1D) array placed inside the waveguide can effectively approximate the normal-incidence response of a 2D periodic structure. The equivalent LC parameters of the mushroom structure, together with the detailed design parameters and electrical characterization of the fabricated PPW fixture, are provided in Supplementary Note~3.

The experimental measurement setup constructed around the PPW is comprehensively illustrated in Fig.~\ref{fig:prototype_setup}d. To characterize the dynamic scattering response, a synchronized dual-channel arbitrary waveform generator (AWG, Ceyear 1652B) is employed to simultaneously provide the broadband test pulse and the low-frequency modulation signal.

For the test signal path, the Gaussian pulse is generated by the AWG with a carrier frequency of $f_{\mathrm{c}} = 1$~GHz, a pulse width parameter of $\sigma = 0.85$~ns, and a time delay of $t_0 = 30$~ns, operating at a sampling rate of $5$~GSa/s:
\begin{equation}
    \label{eq:inc_pulse}
    E_{\mathrm{inc}}(t) = E_0 \cdot \exp\left(-\frac{(t-t_0)^2}{2\sigma^2}\right) \cdot \cos(2\pi f_{\mathrm{c}} t).
\end{equation}
With these parameters, the effective spectral energy is predominantly concentrated within the frequency range of $0.5$ to $1.5$~GHz. This pulse is injected into the PPW via a broadband directional coupler (Mini-Circuits, ZFDC-10-5-S+) and a tapered waveguide transition. The reflected wave from the time-varying sample is subsequently separated by the directional coupler and recorded by a spectrum analyzer (Ceyear 4052F). The absolute reflection spectrum is calibrated against a reference measurement, which is obtained by replacing the metasurface sample with an equal-sized metallic short-circuit plate. The linear reflection coefficient is then computed from the measured spectral power levels as:
\begin{equation}
    \label{eq:ref_cal}
    |\Gamma_{\mathrm{eff}}(f)| = 10^{\left[P_{\mathrm{sam}}(f) - P_{\mathrm{ref}}(f)\right] / 20},
\end{equation}
where $P_{\mathrm{sam}}(f)$ and $P_{\mathrm{ref}}(f)$ are the measured spectral power in dBm of the sample and the metallic reference plate, respectively.

For the modulation path, the temporal control voltage applied to the metasurface is defined as:
\begin{equation}
    \label{eq:mod_voltage}
    V(t) = V_0 \left[1 + M_{\mathrm{v}} \cos(\omega_{\mathrm{m}} t + \phi)\right].
\end{equation}
To physically realize this driving waveform, the low-frequency sinusoidal signal generated by the AWG is first fed into a power amplifier (Ceyear 3871AG-200) to provide sufficient power, ensuring the required modulation depth $M_{\mathrm{v}}$ is achieved. Subsequently, a bias-tee is utilized to combine this amplified AC signal with an appropriate reverse DC bias ($V_0 = 3.5$~V, RIGOL DP932A). To accurately verify the operating state, a passive probe (RIGOL PVP2350) connected to an oscilloscope (RIGOL DS1202) is used to directly contact the metasurface, monitoring the actual modulation amplitude in real time. The dynamic voltage modulation parameters ($f_{\mathrm{m}} = 112.5$~MHz, $M_{\mathrm{v}} = 0.6$, and $\phi = 1.65\pi$) were determined through genetic algorithm optimization by maximizing the Rozanov integral. The theoretical initial phase $\phi$ is physically implemented by configuring the relative time delay between the two output channels of the synchronized dual-channel AWG, which generate the test pulse and the modulation signal, respectively.

To verify the validity of the fabricated sample and the parallel-plate waveguide test fixture, prior to measuring the scattering response of the time-varying system we first characterized the static reflection response under different reverse DC bias voltages, validating the baseline resonant tuning (detailed in Supplementary Note~4 and Supplementary Fig.~3). We also conducted field-circuit co-simulations in CST Microwave Studio to obtain the scattering response of the time-varying metasurface absorber, incorporating the electromagnetic coupling of the 3D structure and the non-linear SPICE model of the varactor diode (detailed in Supplementary Note~5).

\subsection*{Observation of Floquet Harmonic Generation}
\begin{figure*}[!htbp]
    \centering
    \includegraphics[width=\textwidth]{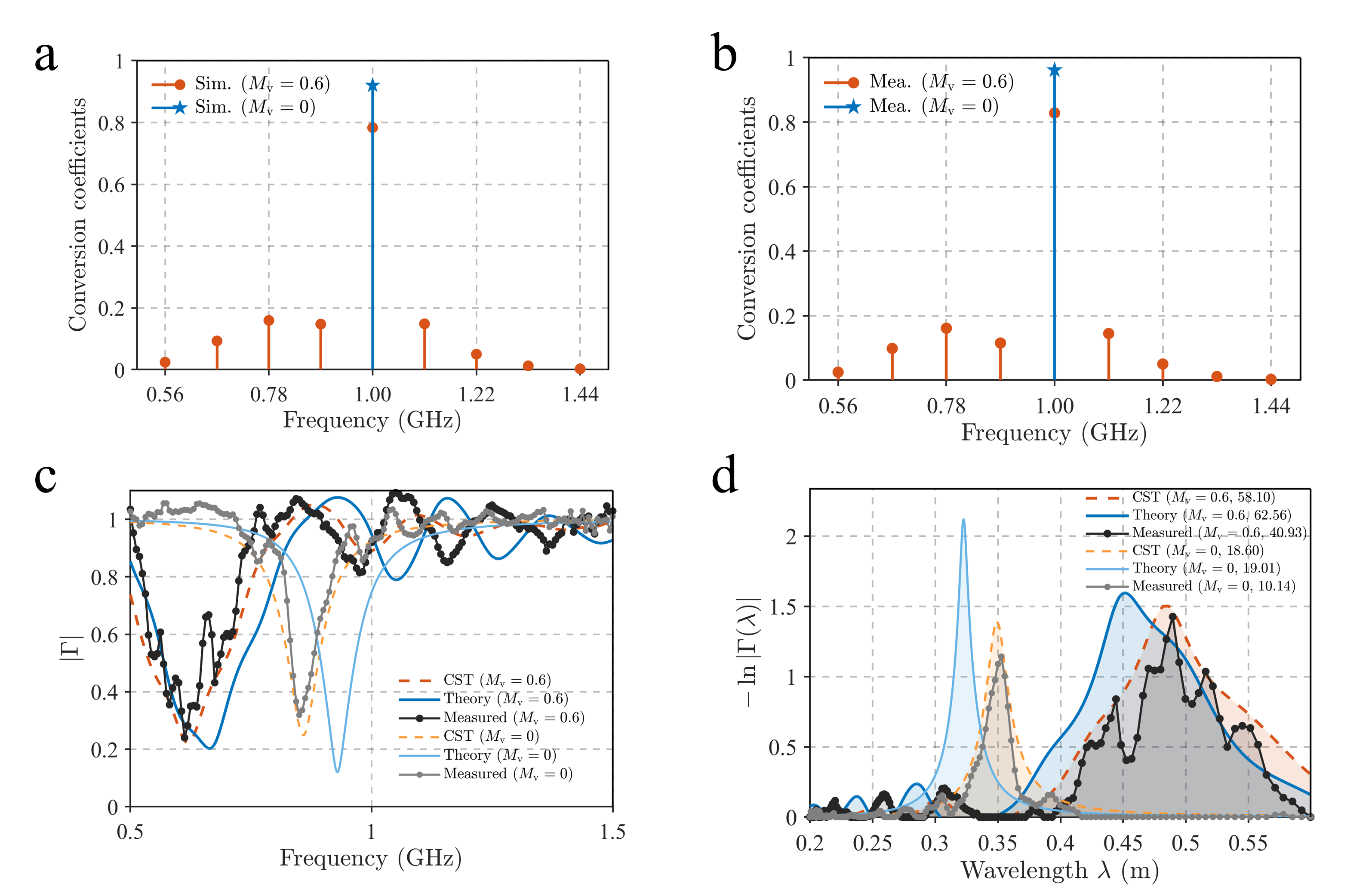}
    \caption{\textbf{Single-frequency and pulsed scattering responses of the time-varying metasurface absorber obtained from CST simulations and experiments.} \textbf{a,b}, Reflection and harmonic conversion coefficients under single-tone excitation at $f_{\text{in}} = 1$~GHz with modulation at $f_{\mathrm{m}} = 110$~MHz: (\textbf{a}) CST field-circuit co-simulation result, and (\textbf{b}) experimental measurement result. \textbf{c}, Comparison of effective reflection coefficients $|\Gamma_{\mathrm{eff}}|$ between static ($M_{\mathrm{v}}=0$) and time-modulated ($M_{\mathrm{v}}=0.6$) states obtained via analytical theory, CST field-circuit co-simulations, and PPW measurements under broadband Gaussian pulse excitation. \textbf{d}, Spectral distribution of $-\ln|\Gamma_{\mathrm{eff}}(\lambda)|$ over the 0.5--1.5~GHz band for the static ($M_{\mathrm{v}}=0$) and modulated ($M_{\mathrm{v}}=0.6$) states, obtained from theoretical analysis, CST field-circuit co-simulation, and measurement. The thickness-normalized Rozanov integral $\mathcal{I}$, evaluated as the area under each curve divided by $d$, is listed in the legend for each case.}
    \label{fig:results}
\end{figure*}

To observe harmonic generation, a single-frequency continuous wave (CW) excitation at $f_{\text{in}} = 1$~GHz is applied. The metasurface is biased at $3.5$~V and modulated by a sinusoidal voltage with $f_{\mathrm{m}} = 110$~MHz and $M_{\mathrm{v}} = 0.6$. Fig.~\ref{fig:results}a,b displays the reflection and harmonic conversion coefficients, calibrated by replacing the metasurface sample with an equal-sized metallic plate. The time-modulated metasurface generates a frequency comb, with scattered energy distributed at the fundamental frequency $f_{\text{in}}$ and harmonics symmetrically spaced in frequency $f_{\text{in}} \pm nf_{\mathrm{m}}$ ($n = 1, 2, 3\ldots$). The measured harmonic amplitudes show reasonable agreement with the CST field-circuit co-simulation predictions.

\subsection*{Metasurface Absorption Beyond the Rozanov Bound}
Under the wideband Gaussian pulse excitation (0.5--1.5~GHz), optimized dynamic voltage modulation parameters ($f_{\mathrm{m}} = 112.5$~MHz, $M_{\mathrm{v}} = 0.6$), and a DC bias of $V_0 = 3.5$~V, the magnitude of the effective reflection coefficient $\Gamma_{\mathrm{eff}}(\omega)$ is extracted. Fig.~\ref{fig:results}c compares the reflection coefficients obtained from analytical theory, CST field-circuit co-simulation, and experimental measurement for both the static ($M_{\mathrm{v}} = 0$) and time-modulated ($M_{\mathrm{v}} = 0.6$) states. Under temporal modulation, the incident wave is coupled to multiple Floquet harmonics, and their coherent interference substantially suppresses the reflection coefficient below 0.9~GHz, resulting in a markedly broadened effective absorption bandwidth.

Using the extracted reflection data under the modulated state ($M_{\mathrm{v}} = 0.6$), the thickness-normalized Rozanov integrals are calculated and visualized in Fig.~\ref{fig:results}d. The integrals derived from the theoretical model, CST co-simulation, and experiment yield $62.56$, $58.10$, and $40.93$, respectively, all of which substantially surpass the fundamental LTI limit of $2\pi^2 \approx 19.74$ for non-magnetic absorbers. The theoretical and simulated values exceed the static limit by factors of approximately $3.2$ and $2.9$, respectively, while the measured value exhibits a certain deviation, which is primarily attributed to the simplifications inherent in the equivalent transmission-line model, the frequency-dependent behavior of the varactor diodes, the fabrication tolerances of the PCB and waveguide assembly, and the finite-size boundary effects of the one-dimensional prototype. Nevertheless, the measured Rozanov integral remains more than twice the static limit, fully confirming that the proposed periodically time-modulated absorber surpasses the Rozanov limit.

\section*{Discussion}

We have experimentally demonstrated, for the first time, a time-modulated metasurface absorber that surpasses the conventional Rozanov bound. By implementing temporal modulation in a spatially structured electromagnetic platform, our results establish a practical pathway toward overcoming the bandwidth–thickness constraints of conventional absorbers. The use of varactor-based capacitance modulation enables continuous electrical control and offers promising opportunities for high-speed modulation and extension toward millimeter-wave frequencies. Moreover, our theoretical analysis reveals that capacitive modulation can generate strong Floquet harmonic coupling at relatively low modulation depths, facilitating broadband spectral redistribution and reflection suppression.

Looking ahead, the flexibility of electrical modulation opens new opportunities for broadband and adaptive electromagnetic absorption. Future developments incorporating real-time feedback control could dynamically optimize the modulation parameters according to the incident waveform, enabling waveform-selective absorption and enhanced broadband performance in changing electromagnetic environments, with potential applications in adaptive electromagnetic shielding and interference suppression.
\section*{Data Availability}
The data that support the plots and findings of this study are available from the corresponding authors upon reasonable request.

\bibliographystyle{naturemag}
\bibliography{reference}

\section*{Acknowledgements}
This work was supported by the National Natural Science Foundation of China (Grant Nos. 12674385 and 62541116) and the Fundamental Research Funds for the Central Universities (Grant No. 3072026RL2501).

\section*{Author Contributions}
X.W. conceived the idea and supervised the project. M.W. carried out the theoretical analysis, numerical simulations, metasurface design, and experimental measurements. Z.L. and M.S.M. contributed to the discussion of the results and provided critical revisions and feedback on the manuscript. All authors contributed to writing and finalizing the manuscript.

\section*{Competing Interests}
The authors declare no competing interests.

\end{document}